\documentclass[twocolumn,twoside,slac_two]{revtex4}
\usepackage{graphicx}
\usepackage{fancyhdr}
\begin{document}

\title{The Extreme Gamma-Ray Blazar S5 0716+714: Jet Conditions from Radio-Band Variability and Radiative Transfer Modeling}

%

\author{M.F. Aller, P.A. Hughes, H.D. Aller,}
\affiliation{University of Michigan, Ann Arbor, MI 48109 USA}
\author{S.G. Jorstad, A.P. Marscher, V. Bala}
\affiliation{Institute for Astrophysical Research, Boston University, Boston, MA 02215 USA}

\author{T. Hovatta}
\affiliation{Aalto University Mets\"ahovi Radio Observatory, Mets\"ahovintie 114, 02540 Kylm\"al\"a, Finland}

\begin{abstract}
As part of a program to identify the physical conditions in the jets of $\gamma$-ray-flaring blazars detected by {\it Fermi}, including the role of shocks in the production of high-energy flaring, we obtained 4 years of 3-frequency, centimeter-band total flux density and linear polarization monitoring observations of the radio-bright blazar S5 0716+714 with the University of Michigan 26-m paraboloid. Light curves constructed from these data exhibit a series of rapid, high-amplitude, centimeter-band total flux density outbursts, and changes in the linear polarization consistent with the passage of shocks during the $\gamma$-ray flaring. The observed spectral evolution of the radio-band flares, in combination with radiative transfer simulations incorporating propagating shocks, was used to constrain the shock and jet flow conditions in the parsec-scale regions of the jet.  Eight forward-moving, transverse shocks with unusually-strong shock compression factors, a very fast Lorentz factor of the shocks of 77,  a bulk Lorentz factor of the flow of 20,  a viewing angle of 12$^{\circ}$,  and an intrinsic opening angle of the radio jet of 5.2$^{\circ}$ were identified.
\end{abstract}

\maketitle

\thispagestyle{fancy}


\section{Overview}
Recent work to localize the site of the GeV emission from blazars using radio-band imaging data has identified a close temporal correlation between activity in the 43 GHz core, a physical region associated with a standing shock \cite{1}, and flaring in the $\gamma$-ray band \cite{2}, while evidence for enhanced centimeter-band activity during $\gamma$-ray flaring has been found for large source samples \cite{3}. Such results support the notion of a common disturbance for the production of both the radio-band and $\gamma$-ray activity and  the localization of the $\gamma$-ray flaring in the parsec-scale region of the jet during at least some flares, e.g. \cite{4}. Hence radio-band linear polarization and total flux density data can be used to probe the  physical conditions at or near to the $\gamma$-ray emission site under the assumption that  $\gamma$-ray flaring originates in the parsec-scale jet.

 The role of shocks in the production of optical-to-radio-band flaring has been widely accepted since the 1980s, and it has been a commonly-cited mechanism for particle acceleration to $\gamma$-ray energies in recent work \cite{5}. However, few detailed studies have been carried out to identify the presence of shocks during $\gamma$-ray flares and to determine their role in the production of $\gamma$-ray flaring. Comparison of simulated light curves based on radiative transfer calculations incorporating the shock paradigm with radio-band total flux density and linear polarization variability observations can be used to identify the properties of shocks (strength, orientation, and sense) and to constrain jet flow conditions. The latter include the bulk motion of the flow, the viewing angle of the jet, the intrinsic jet opening angle, and the degree of order of the magnetic field.  While time-intensive, modeling has the advantage of disentangling complex effects, including relativistic aberration and Doppler boosts, and is preferable to indirect methods based on simple assumptions and the combining of unmatched properties. While useful in obtaining statistical results to delineate parameter space, such procedures can lead to erroneous results for individual sources, in particular where the flow conditions are extreme (fast) and the shocks are strong. 

The intermediate-spectral-peaked (ISP) object S5 0716+714 is both radio and $\gamma$-ray bright, and it has exhibited a history of intense variability across the spectrum, 
including detection in the TeV band, making it well-suited for application of  the shock paradigm. To attain this goal, we carried out intensive monitoring of the linear polarization and total flux density with the University of Michigan 26-m telescope (hereafter UMRAO) at three frequencies (14.5, 8, and 4.8 GHz) during 2008.5 through 2012.5. These data, complemented by millimeter VLBA imaging data at 43 GHz, which probes structural changes in the inner jet of this highly core-dominated source, are modeled here.

\section{The Data}
\subsection{Historical Variability at Radio Band and at GeV Energies}
\begin{figure}
\includegraphics[width=65mm,angle=-90,bb= 73 36 576 720]{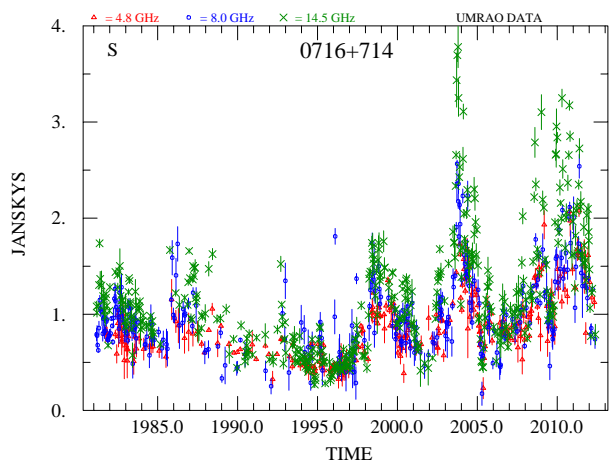}
\caption{Multi-decade total flux density light curve for 0716+714.  The data at 14.5, 8, and 4.8 GHz are denoted by crosses, circles, and triangles respectively.}
\label{f1}
\end{figure}

Centimeter-band total flux density (S) and linear polarization (LP) observations of the $\gamma$-ray-bright BL Lacertae object 0716+714 were obtained from the early 1980s through 2012.5 as part of the University of Michigan (UMRAO) variability program. Such long-term data are useful for placing the observations during the {\it Fermi} era in context. Two-week-averaged long-term total flux density measurements, shown in Figure~\ref{f1}, illustrate nearly continuous activity over 3.5 decades of monitoring. The amplitude of the total flux density exhibited a minimum  near 0.3 Jy in the mid 1990s during the operation of EGRET aboard the Compton Gamma Ray Observatory (CGRO). The source was within the EGRET field of view several times in the 1990s, and 5 detections with
 $\sqrt{TS}\geq$3.5 ($\geq$3$\sigma$ determinations) are listed in the Third EGRET Catalog \cite{6}. These occurred at Viewing Periods with midpoints 1992.05, 1992.19, 1992.46, 1993.55, \& 1995.14. The peak centimeter-band total flux density amplitudes  were considerably lower than those in the {\it Fermi} era, and these EGRET detections occurred during centimeter-band outburst phases which ranged from onset to radio-peak. The characteristic behavior of the centimeter-band variability is notably different after 2002. Thereafter high-amplitude, rapid flaring commenced which has been particularly intense and sustained since the launch of {\it Fermi}.
\subsection{Variability Since the Launch of Fermi}
A blow-up of the UMRAO total flux density and linear polarization data since mid-2008 and the weekly-binned $\gamma$-ray photon flux light curve in the 0.1--200 GeV band is shown in Figure 2. The photon fluxes (panel 1) were obtained using {\bf ScienceTools}--v9r27p1 and P7SOURCE\_V6 event selection. The LAT data were extracted within a 10$^\circ$ region of interest (ROI) centered upon the position of the target. These used an unbinned likelihood analysis (tool gtlike) to determine the photon fluxes by including in the model all of the sources within 15$^{\circ}$ of the target and by freezing the spectral index of all sources to the value in the 2FGL catalogue. Daily-averaged  UMRAO centimeter-band data are shown in panels 2-4. With the higher sampling attained since the launch of {\it Fermi}, the individual radio-band flares comprising an outburst are resolved during each activity phase; sufficient resolution in the data is an important requirement for identifying the number of individual flares within an outburst envelope and is used to set the number of shocks. The
linear polarization is shown in panels 3 and 4 in the form of fractional linear polarization and electric-vector position angle (EVPA). There is a 180$^{\circ}$ ambiguity in the determination of the UMRAO EVPAs, and our convention is to restrict the UMRAO EVPA values to lie in the range of 0$^{\circ}$ - 180$^{\circ}$. In this plot, however,  we have allowed the range to be slightly larger to minimize the occurrence of apparent jumps. 

VLBA imaging data provide important additional constraints in the analysis of the variability in this source. Source-integrated 15 GHz data from the MOJAVE website (http://www.physics.purdue.edu: orange squares) and core fluxes obtained from the BU blazar program, VLBA-BU-BLAZAR, (http://www.bu.edu/blazars/VLBAproject.html) providing monthly data at 43 GHz are included in this figure. The agreement between the 15 GHz MOJAVE imaging data and the 14.5 GHz source-integrated UMRAO measurements confirms that there is no significant contribution from extended VLA-scale structure to either the total or polarized flux.  The redshift for this source is not known directly from spectroscopic measurements. However, adopting a value of 0.3 has led to a maximum apparent component speed of $\geq$40c based on the analysis of 5 moving MOJAVE components, and  a change in the 15 GHz projected inner jet position angle in late 2009-early 2010 is identified from the structural changes with time \cite{7}. A  preliminary analysis of the 43 GHz VLBA monitoring data obtained in the Boston University Blazar program identified complex structural changes in the inner jet region and both stationary and moving jet components; 3 of the component ejections (times at which the feature separated from the core) were temporally-associated with $\gamma$-ray flares during the time window presented \cite{8}. Comparison of the amplitudes and variations in the millimeter and centimeter band gives a measure of the opacity between the respective emission sites and a fiducial location in the jet flow since the 43 GHz `core' is associated with a physical feature in the flow. The fact that both the linear polarization and the total flux density measurements characteristically track at 15 and 43 GHz suggests that either the 43-15 GHz emission region is optically thin during this time window, or that the emission sites are spatially close to each other.

Cross-correlations of the fluxes from ground-based and satellite instruments at a variety of wavebands have been carried out for segments of the data sets in a number of recent papers, e.g. \cite{9}, and characteristic time scales identified using structure functions and periodograms \cite{10}. This work identifies complex relationships which are overall consistent with the production of flaring by a common disturbance which propagates outward in the jet with time. The matching of individual total flux density flares in the radio and $\gamma$-ray light curves, however, is complicated by the presence of nearly continuous activity in the radio-band in the post-launch time-period, intrinsic difference in the  doubling time for flaring in the two bands, and by geometric effects, e.g. \cite{11}, including changes in jet orientation with time and jet curvature.
\begin{figure}
\centering
\includegraphics[width=65mm]{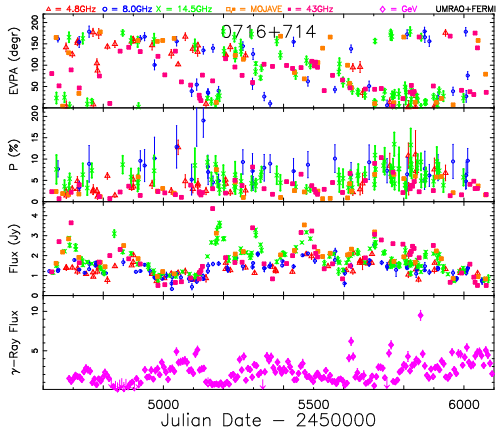}
\caption{The $\gamma$-ray and centimeter-band total flux density and linear polarization light curves: May 13, 2008 through June 12, 2012. From bottom to top: (panel 1) the weekly-binned $\gamma$-ray light curve in units of photons/s/cm$^2$ x 10$^{-7}$,  and (panels 2-4) radio-band total flux density, fractional linear polarization (P\%), and electric vector position angle (EVPA) shown as daily-averaged data. Core fluxes at 43 GHz from the Boston University program (magenta squares) and  source-integrated 15 GHz MOJAVE VLBA data (orange squares) are included.} \label{f2}
\end{figure}

\section{Radiative Transfer Modeling}
The framework and assumptions adopted in the radiative transfer modeling are detailed in \cite{12}. We adopt a scenario in which propagating shocks pass through a region of the jet containing a predominantly-turbulent, passive magnetic field; the support for the presence of a turbulent magnetic field comes from the UMRAO result for hundreds of sources showing low levels of fractional polarization (typically a few percent) on timescales of decades, and no evidence for the high values (P$>$20\%) of the fractional linear polarization in the parsec-scale domain predicted by some models. Such high fractional linear polarizations might be expected for isolated regions of highly-ordered magnetic field, e.g. as found in some jet components from VLBI polarimetry observations, but not in the underlying quiescent flow. While the magnetic field is predominantly turbulent in our model, an additional ordered axial magnetic field component was initially included in the modeling  to reproduce the stable EVPAs observed in UMRAO sources during the relatively-rare time windows of quiescence, but early modeling revealed that this axial magnetic field can have a significant effect on the simulated light curves and that higher values than the initially-adopted value of 2\% in magnetic energy were required to match the data. The value of the axial magnetic field is one of the parameters determined by the modeling. The
shocks compress the plasma in the emission region, increasing the emissivity and producing an increase in the degree of order of the magnetic field. The signature of the passage of a shock in the light curves is an outburst in total flux density, an increase in the  fractional linear polarization, and an ordered swing in the EVPAs, and it is the spectral evolution with time in both the linear polarization and the total flux density which constrains our models. The number of shocks used in the simulation is established from the structure apparent in the total flux density and linear polarization light curves, combined with the expected burst profile for a single shock \cite{12}.  The
shocks are allowed to be oriented at an arbitrary direction relative to the jet flow and are specified by two angles.  The angle $\eta$, the shock obliquity, specifies the angle relative  to the direction of the upstream flow. A second angle specifies the azimuthal direction of the shock normal, but as discussed in \cite{12}, the simulation is relatively insensitive to the choice of the value for this free parameter. For simplification it is assumed that the shock occupies the cross section of the flow and propagates at a constant rate.The attributes which specify a shock are its onset time, length, compression factor, sense (forward or reverse), and orientation. 
\begin{table}[t]
\begin{center}
\caption{Shock Parameters}
\begin{tabular}{|l|c|c|c|c|c|c|c|c|}
\hline  \textbf{Shock}   & 1     & 2     & 3     &   4     &  5     &  6     &  7     &  8   \\
\hline T$_{o}$-2000  & 0.85  & 08.9  & 09.8  &  10.15  &  10.6  &  11.2  &  11.4  &  11.8  \\
\hline $\kappa$ & 0.18  & 0.18  &  0.18 &   0.2   &  0.17  &  0.2   &  0.27  &  0.25  \\
\hline
\end{tabular}
\label{t1}
\end{center}
\end{table}

\section{Modeling Results for 0716+714: Shock and Flow Parameters}
The iterative procedure used for the analysis  is illustrated in \cite{13} which presents details and results for  events modeled in three additional $\gamma$-ray-bright blazars, and the sensitivity of the model to changes in the values for the key model parameters is examined in \cite{14}. 
An initial shock obliquity was chosen based on the changes in the UMRAO EVPA light curves. In this source swings through approximately 90$^{\circ}$ occurred, consistent with the passage of a shock whose front is transverse to the jet axis. By assumption  all of the shocks during the activity modeled have the same orientation.  The length of each shock (defined as the evolved extent of the shocked flow) was set at 0.005 times the length of the flow. The shock sense was found to be forward moving; we explored the possibility of reverse  shocks, but these led to inconsistencies with  results based on VLBI measurements.
The individual start times and compression factors ($\kappa$) required to reproduce the spectral and temporal variability are listed in Table~\ref{t1} for each shock. Typical compressions for other $\gamma$-ray-flaring sources are 0.5 - 0.8, and the shocks in 0716+714 require unusually strong compressions as suggested by the very rapid rises in the light curves. 

 Parameters specifying the jet are given in Table~\ref{t2}.  The value of the optically-thin spectral index $\alpha$ (S$\propto\nu^{-\alpha}$) was set to 0.25 based on the rather flat total flux density spectra in general in the UMRAO sources, and the fiducial `thermal' Lorentz factor of the energy spectrum was arbitrarily set to 1000 at 8 GHz. The model assumes a power law distribution of the radiating particles with cutoff $\gamma_i$.  The low energy cutoff, determined to be 50 in this source, was constrained primarily by the EVPAs. The bulk Lorentz factor was determined from the observed P\%; the derived value of 20 is high compared to the values in the range 5-10 determined for our other modeled sources, but is consistent with the high values of $\beta_{app}$ determined from VLBI measurements for the fastest components. The flow viewing angle is determined primarily by the observed  fractional linear polarization but uses the observed EVPA and the range of its change as a secondary constraint. For comparison, typical values found from our modeling of other sources are in the range 1.5$^{\circ}$ to  4$^{\circ}$, and our derived value of this parameter is unusually high for a blazar. The apparent component speed,
 $\beta_{app}$, is computed from the shock Lorentz factor and the viewing angle. The shock Lorentz factor, in turn, comes from the bulk Lorentz factor of the quiescent flow and the shock strength (compression factor). 

\begin{figure*}[t]
\centering
\includegraphics[width=65mm,totalheight=65mm]{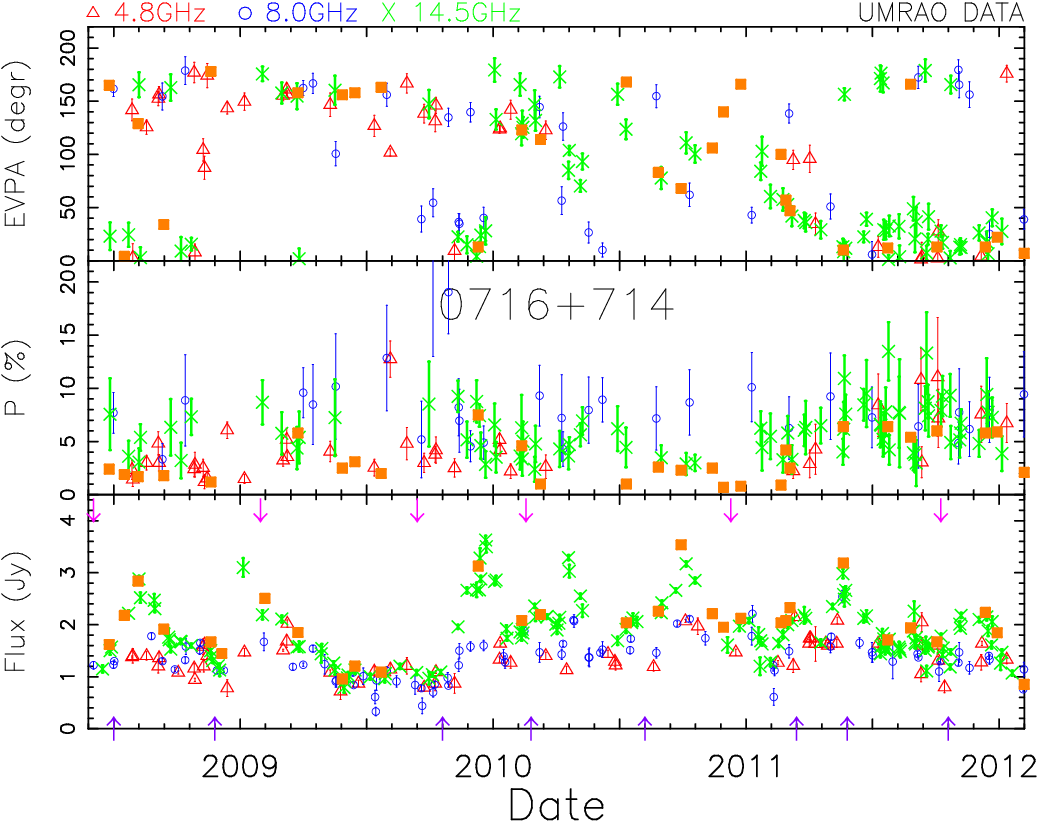}%
\hspace{0.2in}%
\includegraphics[width=65mm,totalheight=64mm]{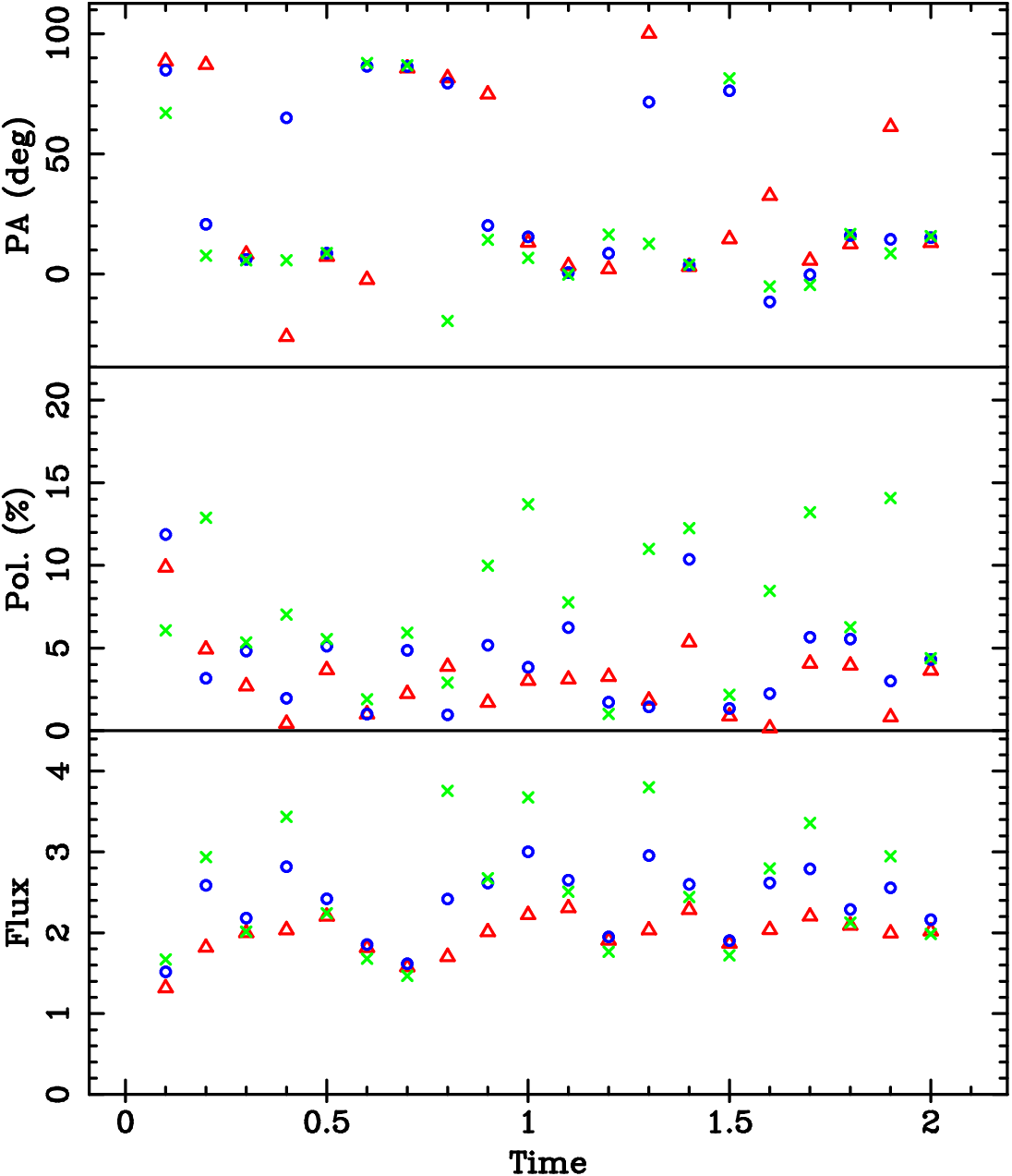}
\caption{Left: Observed total flux density and linear polarization for the time segment modeled. The symbol convention follows Figure~\ref{f1}. The MOJAVE source integrated values are included for comparison.  In panel 2 the error bars are omitted for clarity. The upward arrows along the abscissa of the lower panel (Flux) mark the shock start times. The downward arrows along the top of this panel mark the component separation times from new model-fitting of the 43 GHz VLBA data. Right: Simulated total flux density and linear polarization light curves based on the shock and jet  parameters given in the text. Time is expressed in arbitrary units using 20 time steps over the time window modeled. The three frequencies are color and symbol coded to match the convention adopted for displaying the data.} \label{f3}
\end{figure*}

 The simulated light curves based on the shock and jet parameters given in the tables are shown in Figure~\ref{f3} right and the UMRAO data used as constraints are shown in Figure~\ref{f3} left. The scaling of time  in the model light curves is set by the duration of the activity modeled, while the total flux density is scaled to match the peak amplitude of the total flux density at the highest UMRAO frequency, 14.5 GHz. Recall that there are 180$^{\circ}$ ambiguities in the determination of the EVPAs. Hence in the comparison of the data and the model only the range of the swings should be considered and not the values. The shock onsets are marked along the abscissa of the lower panel of the left plot. These mark the times at which the leading edge of the shock enters the flow, and they do not correspond to times at which the brightness centroids of new VLBI components cross those of the stationary `core'. Those times, based on the 43 GHz data, are also indicated in the figure.

\subsection{ Deviations between the Model and the Data}
 While the model is able to reproduce the general character of the variability, including the spectral behavior as a function of time, the amplitude range of the total flux density flares,  and the global event shape and the position of features, there are some important differences between the observed and simulated light curves which indicate that refinements to the model are required. Geometric effects have been cited in a number of papers, and an association between changes in the inner jet PA at 43 GHz and with  $\gamma$-ray activity is proposed in \cite{15}. These would be expected to affect the character of the total flux density light curves which are successfully reproduced in terms of the  amplitude range and the spectral behavior with time.  More difficult to account for are the differences  between the observed and simulated fractional linear polarization.  Refinements to the model are required to reproduce the spectral character of the 8 GHz polarimetry data, especially during shocks 2 through 5.  Further, in the early part of the simulation, the values of P\% are too high at all three frequencies (e.g. at 4.8 GHz 10 versus 5). Modifications which we hope to explore in future work in an effort to improve the fit  of the model include  allowing for a range of shock obliquities in the simulation and including a modest contribution from an additional ordered (possibly helical) magnetic field component.

\section{Discussion and  Conclusions}
 Radiative transfer modeling of the UMRAO data incorporating 8 forward-moving shocks is able to reproduce the primary features of the spectral variability in both total flux density and linear polarization in 0716+714. The modeling identifies a high bulk Lorentz factor of the flow, consistent with prior results, a wide intrinsic jet opening angle compared with other blazars \cite{13} and unusually strong shock compressions. The apparent speeds of the emission pattern are high, but less than the Lorentz factor since the observer lies outside of the critical cone of the fast flow. The viewing angle, a parameter which is very well constrained by both the linear polarization and the total flux density measurements, is higher than found for the outbursts which we modeled in other $\gamma$-ray-bright sources. These results are consistent with the impression from the light curves alone that the jet conditions in this source are extreme. 

The fact that the 43 GHz VLBA and 14.5 GHz monitoring data in {\it both} the linear polarization and the total flux density track so well during the $\gamma$-ray flaring is unusual compared to other blazars. The modeling identifies that the radio-band emission originates in a partially optically-thick part of the jet, and this result rules out the optically thin scenario suggested as a possible  explanation  to explain the tracking of the millimeter and centimeter-band flux and linear polarization data. The millimeter and centimeter-band  data, combined with the modeling, support an interpretation of spatially-close centimeter and millimeter-band emission sites. The localization of the $\gamma$-ray site relative to the 43 GHz core region remains a controversial issue. The variability in this source is complex, and the broadband data exhibit time-dependent behavior which complicates attempts to establish significant correlations. These may arise from a complex mix of changing physical conditions within the jet, geometric effects, and more than one $\gamma$-ray emission site. The correlations identified are consistent with the production of at least some $\gamma$-ray flares within the parsec-scale radio jet, and the modeling provides relevant source parameters for those $\gamma$-ray flare events. 

 This work illustrates the importance of multifrequency linear polarization monitoring data. They  have the power to directly probe the magnetic field direction and degree of order in blazars and to provide relativistic jet flow and shock properties which cannot be obtained directly from observations.

%


\begin{table}[t]
\begin{center}
\caption{Jet Parameters}
\begin{tabular}{|l|r|}
\hline \textbf{PARAMETER} & \textbf{VALUE}
\\
\hline Spectral Index          &  0.25   \\
\hline Fiducial Lorentz Factor & 1000  \\
\hline Cutoff Lorentz Factor   & 50 \\
\hline  Bulk Lorentz Factor    & 20\\
\hline Jet Opening Angle       & 5.2$^{\circ}$   \\
\hline Viewing Angle           & 12.0$^{\circ}$   \\
\hline  Shock Lorentz Factor   & 77               \\
\hline Shock $\beta_{app}$     &  9.5c     \\
\hline Energy in the Axial {\bf B} Field & 36\%  \\
\hline
\end{tabular}
\label{t2}
\end{center}
\end{table}

\bigskip 
\begin{acknowledgments}
Funding was provided by NSF grant AST-0607523 and NASA {\it Fermi} GI grants NNX09AU16G, NNX10AP16G, and NNX11AO13G (U. Michigan); NSF grant AST-0907893 and NASA {\it Fermi} GI grants NNX08AV65G and NNX11AQ03G (Boston U.); and Academy of Finland project number 267324 (T.H). Computational resources and services were provided by Advanced Research Computing at the University of Michigan, Ann Arbor. This project has made use of data from the MOJAVE website which is maintained by the MOJAVE team.
\end{acknowledgments}

\bigskip 

\end{document}